\documentstyle[prl,epsfig,twocolumn,aps]{revtex}
\begin{document}
\twocolumn[\hsize\textwidth\columnwidth\hsize\csname@twocolumnfalse%
\endcsname
\title{A Minimal Growth Equation For Conservative Non
Equilibrium 
 Growth From Vapor}  
\author{S.V. Ghaisas} 
\address{
 Department of Electronic Science, University of
Pune, Pune 411007,
India}
 
\date{\today}
                  
\maketitle 
 
\begin{abstract}  
A continuum growth equation in 1+1 dimensions is obtained by considering
the 
contributions to the underlying current from kinetics of adatoms  
on the surface 
. These considerations
reproduce the slope dependent  
term that represents a stable or an unstable growth depending 
 on the sign of the associated coefficient.  
 In addition an $h\rightarrow
-h$ 
symmetry breaking nonlinear term is obtained. Implications 
of this term on the stable$\rightarrow$ unstable growth 
transition are discussed.   
This 
term is tilt independent and is expected to be present
independent 
of the source of the current in the conserved growth.
computer simulation based on a zero diffusion bias model 
 is used to verify some of the predictions of the minimal 
growth equation. 
It is shown that the growth equations  
for cellular automata type models involving instantaneous
relaxations can be obtained from the proposed method by 
addressing the corresponding kinetics of adatoms.

\end{abstract} 
 
\pacs{PACS numbers: 
68., 68.55.-a,81.15.-z, 68.55.Ac}                            
] 
\narrowtext
 A continuum description of conserved growth from
vapor is obtained 
from Langevin type equation $\partial_{t}h+{\bf \nabla
 \cdot J}=F$ where 
$h({\bf r},t)$ is height function , {\bf J} is current
due to adatom 
relaxation on the growing surface and $F$ is the 
average flux with 
 white noise. The current {\bf J} includes an
equilibrium and 
a nonequilibrium(NE) contribution. Most of the
experimental growth 
is far from equilibrium and has been studied over last
many years
\cite{bar,kr1}. Description of such growth is essentially through 
the nonequilibrium current {\bf J$_{NE}$}. The contributing 
terms to {\bf J$_{NE}$} are obtained by comparing the behaviour 
of growth equation containing sets of different terms, 
satisfying required symmetry and invariances of the problem, 
 with the experimental 
observations\cite{len1,ld} or, appealing to the underlying kinetics 
\cite{vill1,vill2}. Former is an empirical approach while later,
 a direct approach. It is desired that from kinetic considerations 
, {\bf J$_{NE}$} could be constructed, describing the experimental 
observations.Such an approach was adopted by Villain\cite{vill2} 
earlier. In reference\cite{vill2}, from computer based solutions 
of the equations describing step velocities in 1+1 dimensions, 
and the consequent interface profiles, it was concluded that 
 the local form of the growth
equation should comprise
of a) a term of the form $\frac {\pm
m}{(1+a' |m|)(1+b' |m|)}$ where $a'$ 
and $b'$ are constants, related to the diffusion
kinetics, b) an 
$h\rightarrow -h$ symmetry breaking term and c) a term
of the form 
$\nabla^{4}h$. 
   
  In what follows we show that these terms can be {\it derived} 
by appropriately 
identifying the kinetic processes that contribute to the
nonequilibrium current. To our knowledge the method proposed 
in this work has not been used to obtain {\bf J$_{NE}$} in 
any of the previous work related to growth from vapor. 
Most of our discussion will be pivoted around the
symmetry breaking term 
that assumes the form $\nabla^{2}(\nabla h)^{2}$ for
small inclinations 
while at large slopes takes the form
$\frac{1}{m^{3}}\partial_x{m}$ thus 
unifying the descriptions\cite{kr1,vill2} at small and large slopes. While
obtaining the NE contributions 
to the current, we will consider growth with
Schwoebel-Ehrlich (SE)\cite{se}  
barrier to account the stable as well as unstable
growth. However 
the method of obtaining these contributions can be
applied to any  
other models not employing the SE barrier \cite{tl,sdp}. 
 We have applied 
it later to a stochastic model with cellular automata
type rules that 
mimic the molecular beam epitaxial (MBE) growth at
low temperature
\cite{dt} referred in the literature as DT model.

\begin{figure} 
\epsfxsize=\hsize \epsfysize = 1.0 in
\centerline{\epsfbox{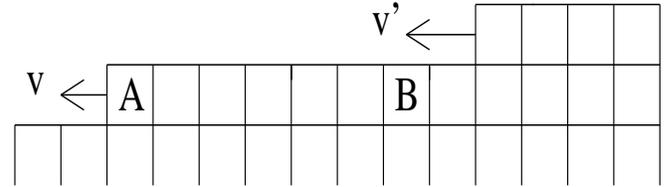}}
\caption{A typical step structure formed during 
growth along positive slope. $v$ and $v'$ are velocities of the steps 
.  
}     
\label{step}  
\end{figure}

 Consider growth on a 1 dimensional flat substrate
with lattice 
constant $a$. We will consider the situation depicted
in Fig. \ref{step}  
for obtaining various contributions to the current where steps 
are such that positive slope is obtained.
Adatoms are randomly
 deposited on the substrate. Let $D_{s}$ be the
diffusion constant 
on the terrace, $l_{c}$ be the average distance that
an adatom travels 
on a terrace before encountering another adatom.
Detachment from steps 
or nuclei on the terrace is negligible. Adatoms
hopping down the descending 
steps contribute to the downward current $j_{d}$ while
those hopping 
on the terrace can get attached to an ascending step
contribute to the 
in-plane current $j_{i}$. Referring to Fig. \ref{step} , the
adatoms reaching 
site $A$ and hopping down the step to the left
constitute $j_{d}$ and 
those reaching site $B$ and hopping to the right
towards the 
ascending step constitute $j_{i}$. The net current is
$j_{d}+j_{i}$. 
These are obtained as follows, 

    $ j_{d(i)}$=(local density of site A(B))$\cdot  $(flux of
adatoms \\
 approaching 
              A(B))$ \cdot $(probability for hopping across
A(B)).
\begin{equation}
\end{equation}   

Since the relaxation of adatoms is through the 
diffusion on the terrace or across the step edges, 
the density of sites A and B is 
same as 
density of steps which is approximately given by
$\frac {|m|}{1+|m|}$, 
where $m$ is the local slope.   

In the absence of nucleation, lateral flux(LF) approaching site B or A is 
$\frac{\pm \hat n F}{2|m|a^{-1}}$, $a|m|^{-1}$ being the average local 
terrace width and  $\hat n$ is unit vector in the x direction.  
When terrace size is large enough, this flux is restricted due to 
the nucleation. The nucleation process will restrict the diffusion 
to  an average  length $l_{c}$ on a 
large terrace. As a result, even for very large terraces, the 
approaching flux LF is almost constant. The effect of nucleation 
is incorporated by introducing $l_{c}$ in the expression for 
LF as $\frac{\hat n F}{2(l_{c}^{-1}+|m|a^{-1})}$, so 
that for small slopes the expression is reduced to a constant value.
Let $P_{A}$ and $P_{B}$ represent the probabilities of hopping across 
sites A and B respectively. The Schwoebel length \cite{vill1} $l_{s}
\propto (P_{B}-P_{A})$ . Hence from the Eq.(1) the current is 
\begin{equation}  
   {\bf j}_{s}=\frac {\hat n |m|F(P_{B}-P_{A})
}{2(1+|m|)(l^{-1}_{c}+|m|a^{-1})}
\end{equation}

The LF approaching sites A and B is however 
 modified due to the relative motion of the steps. Consider the 
situation in Fig.\ref {step}  where $v$ is the velocity of the step bearing 
the terrace while $v'$ that of the higher one on the positive slope.
For $v'>v$ the terrace width is reduced, depleting the LF approaching 
sites A and B. The reduction in flux is $\propto \delta v$ where 
$\delta v=(v'-v)$. Adatoms hopping across upper step as well as those 
attaching in-plane, both contribute to the velocity of the step. Thus 
the velocity $v\propto j_{s}$ except that the coefficient $(P_{B}-P_{A})$
 is replaced by $(P_{A}+P_{B})$. Hence $\delta v\propto \frac{(P_{A}+P_{B})}
{l^{-1}_{c}+|m|a^{-1}}\partial_{x}\frac {|m|}{2(1+|m|)(l^{-1}_{c}+|m|a^{-1})}
$. Corresponding current {\it will not depend on} $(P_{B}-P_{A})$ since this 
part of the flux is removed from the LF. The current is therefore obtained 
by multiplying the LF by density of steps. 
 $P_{A}$ and $P_{B}$ are relative 
probabilities so that $P_{A}+P_{B}=1$.  
Accounting for this effect the expression for 
the current becomes, 
\begin{eqnarray}  
  {\bf j}(x)&=&\frac {\hat n |m|F (P_{B}-P_{A})}{2(1+|m|)
(l^{-1}_{c}+|m|a^{-1})}
             \nonumber\\
&&-\frac {\hat n |m|F}{(1+|m|)(l^{-1}_{c}+|m|a^{-1})}\nonumber\\
&&\partial_{x}
\left(\frac {|m|}{2(1+|m|)(l^{-1}_{c}+|m|a^{-1})}\right)   
\end{eqnarray}
In the limit of small $m$, the current reduces to, 
\begin {equation} 
 {\bf j}(x)=\hat n (P_{B}-P_{A})|m|l_{c}/2
-\hat n l^{-1}_{c}
\partial_{x}(m^{2}F)
\end {equation}  
The second term in general will have the form ${\bf \nabla}(\nabla h)^{2}$. 
This term was derived using the Burton-Cabrera-Frank (BCF) theory and 
assuming that at small slopes the particle density on the terraces 
depends on the even powers of local gradient \cite{kr1,vill2}. In the 
presnt case the difference in the velocities of two consecutive steps 
on a slope is related to the difference in the adatom densities on their 
terraces. However we do not explicitly assume the 
dependence of density on the even powers of slope. Further, in ref. \cite{ld} 
it was conjectured that such a term can arise due to the differences in 
the velocities of the steps near the top and the bottom of a profile.  
 
 In the limit of large slope, $|m|a^{-1}\gg l^{-1}_{c}$ the current reduces 
to, 
\begin {equation} 
{\bf j}(x)=\frac {\hat n (P_{B}-P_{A})aF}{2|m|}
-\frac {\hat n Fa^{2}}
{2|m|}\partial_{x}\left(\frac {1}{|m|}\right) 
\end {equation}  
The second term is proportional to $\frac {1}{m^{3}}\partial_{x}m$, 
which was derived in the large slope limit in reference \cite{vill2}. 
Thus the geometrical dependence of the symmetry breaking term 
in Eq.(3) exactly matches with previously derived two terms in the 
small and large slope limits. This shows that present method of 
deriving current, appropriately accounts for the physical processes 
contributing to it. 

We further argue that a curvature dependent current must be present 
in any adatom relaxation process that involves downward hops across 
the descending step edges. This argument is based on the observation 
that, in a 1+1 dimensional simulation, 
if adatoms are restricted completely to 
the in-plane hops (infinite SE barrier) then correlations do not 
grow beyond the diffusion length. On the other hand, when such hops
 are allowed, correlation length for stable growth and mound size for 
unstable growth increases in time\cite{svgunpub}. Hence if the downward 
hops are allowed but the current is tilt-independent, then it may be 
expressed as a linear combination  $a_{1}\nabla^{3} h+a_{2}\nabla^{5} h
+...$ including the nonlinear terms of the form ${\bf \nabla}(\nabla^{3} h)
^{2}$. We will retain only $\nabla^{3} h$ in the current corresponding 
to our minimal growth equation. Thus the form of the current corresponding 
to the minimal growth equation is 
\begin {eqnarray}
  {\bf j}(x)&=&\frac {\hat n |m|F(P_{B}-P_{A})}{2(1+|m|)(l^{-1}_{c}+
|m|a^{-1})}\nonumber \\ 
&&-\frac {\hat n|m|F}{2(1+|m|)
(l^{-1}_{c}+ 
|m|a^{-1})}\nonumber \\
&&\partial_{x}\left(\frac {|m|}{(1+|m|)(l^{-1}_{c}+|m|a^{-1})}\right) 
+k\frac{\partial^{3}h}{\partial x^{3}}  
\end {eqnarray} 

The first term has been studied widely as the stable growth mode
\cite{ew,fam} and as the unstable growth mode\cite{len}.    
In order to find the effect of asymmetry term on the growth 
exclusively, we have performed simulations of a 1+1 dimensional 
solid-on-solid model with {\it no diffusion bias}. A fourth ordered 
equation was earlier proposed by Villain \cite{vill1} for similar situation.  
Under this condition 
$P_{A}=P_{B}$ and first term vanishes in the Eq(3). The resultant 
growth equation in the moving frame with growth front is of the form, 
\begin {equation}
\partial_{t}h=\nu\nabla^{4}h+\nu_{a}\partial_{x}\left
(\frac {m}{(1+|m|)((l^{-1}_{c}+
|m|a^{-1})}\right)^{2}+\eta 
\end {equation} 
where, $\nu_{a}$ is the appropriate constant for the asymmetry term 
and $\eta$ is white noise with $<\eta(x',t')\eta(x,t)>=D\delta(x'-x)
\delta(t'-t)$. In the limit of small slopes, 
renormalization group (RG) analysis 
shows that the roughness exponent $\alpha=1$ and the roughness  
evolves  with the exponent $\beta=1/3$\cite{ld}. For large slopes, 
asymmetric term, on power counting leads to $z-3\alpha$, while 
noise term gives $z-1-2\alpha$. If $\nu_{a}$ is assumed to 
be invarient under renormalization, then again $\alpha=1$ and $\beta=1/3$ 
as in the case of small slope.   
Corresponding model is as follows.
Atoms are rained on a 1-d substrate of length $L$ 
randomly with constant flux. On deposition a given adatom is allowed to 
hop n times, as in a random walk. If the hopping adatom encounters 
another adatom before n hops are exausted, the adatom stays there 
permanently. If n hops are exausted without any encounter, it stays 
permanently at the last position occupied after n hops. Thus for 
hopping, all sites with zero neighbor are equivalent, ensuring 
$P_{A}=P_{B}$.     
We have measured width $w_{2}$, 
height-
height correlations $G({\bf r},t)$ and skewness $\sigma$ 
for this model where, $w_{2}=\frac{1}{N}
\sum_{i}(h_{i}-\bar h) \sim t^{2\beta}$ and $G({\bf
r},t)=\frac{1}{N} 
\sum_{{\bf r'}}(h({\bf r}+{\bf r'},t)-h({\bf
r'},t))^{2}$.   
 The skewness
$\sigma=w_{3}/w_{2}^{3/2}$ ,
where $w_{3}=\frac{1}{N}\sum_{i}(h_{i}-\bar
h)^{3}$\cite{kim}. 

\begin{figure} 
\epsfxsize=\hsize \epsfysize= 3.0 in
\centerline{\epsfbox{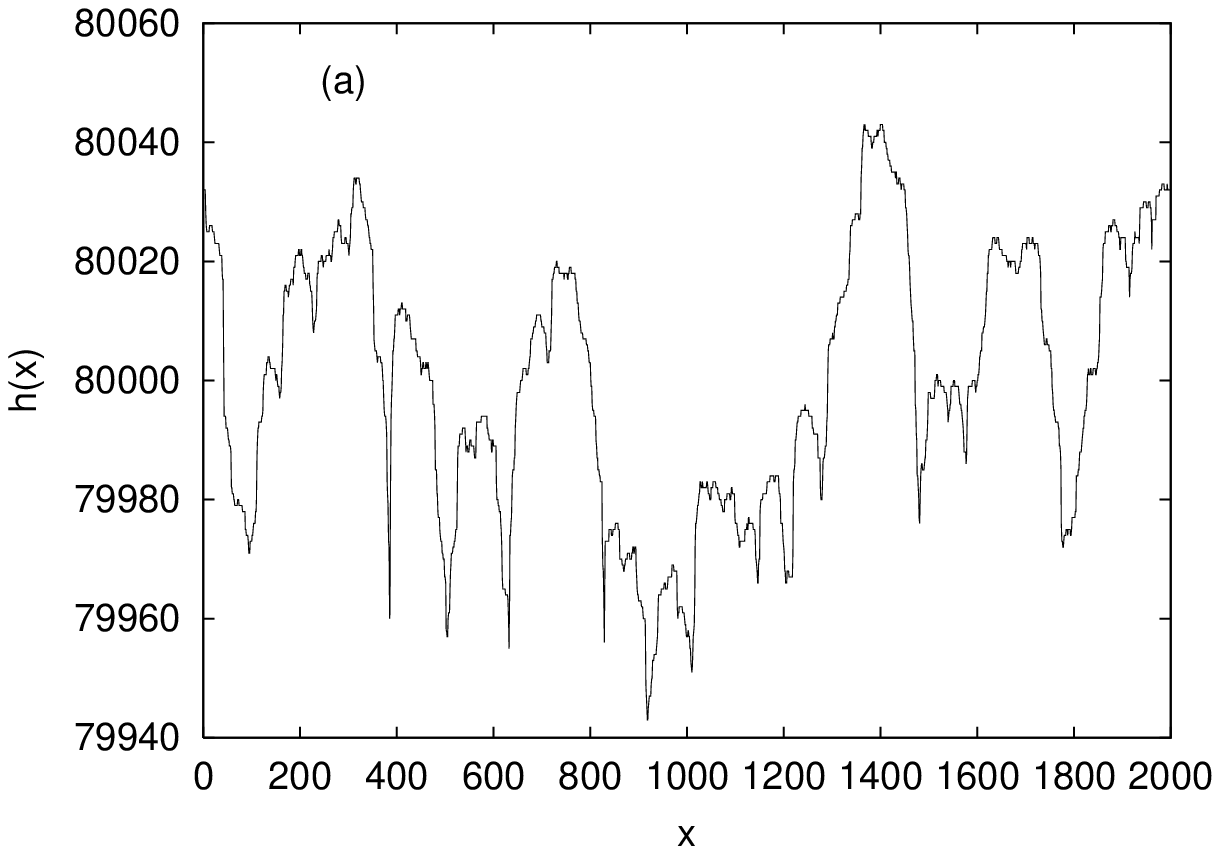}}
\epsfxsize=\hsize \epsfysize = 3.0 in
\centerline{\epsfbox{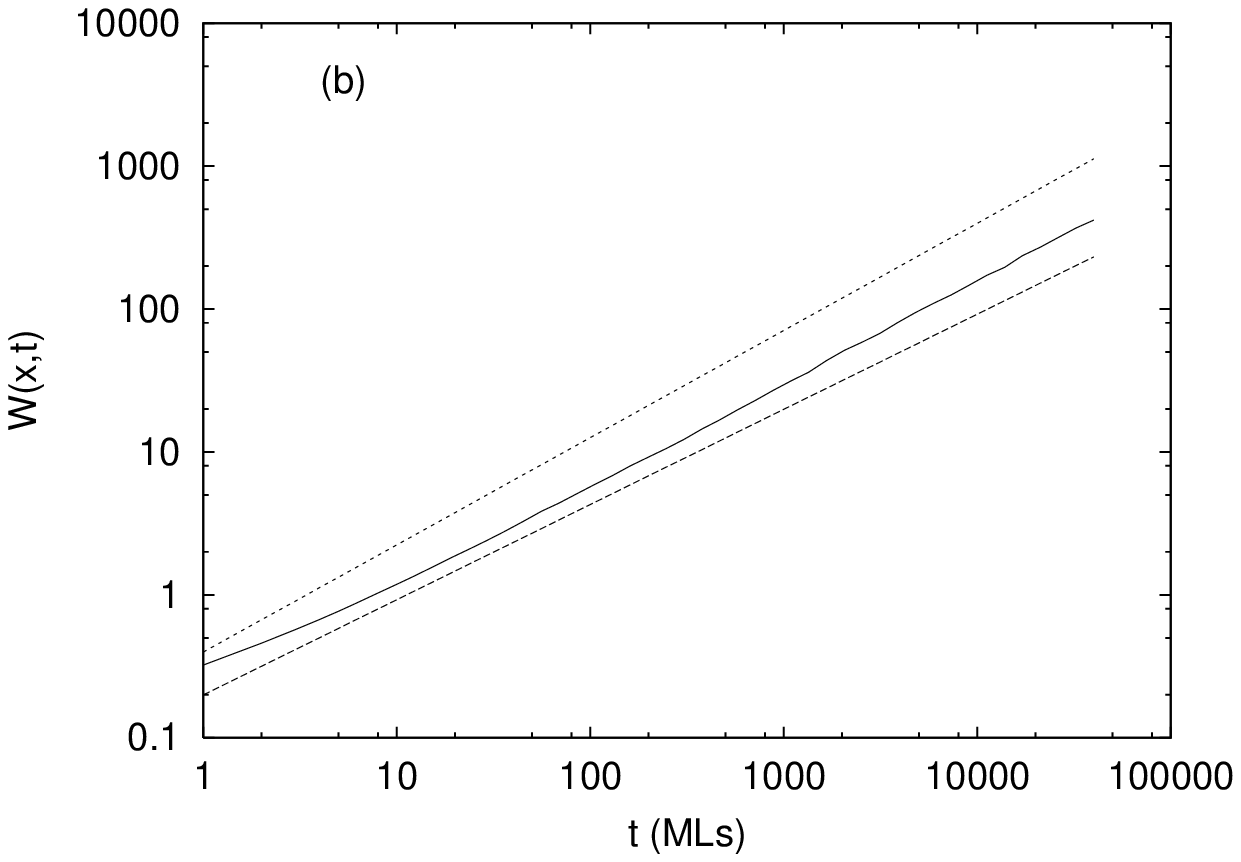}}
\caption{(a) Morphology of the surface after 80000 
number of layers.  
 (b)Plot of width as a function of time. Straight 
lines with slope $3/4$ and $2/3$ are drawn for reference.  
}    
\label{morfw}  
\end{figure}

Presence of $\nabla^{4}h$ term is verified from the flatness of the 
saturated width for small $L$. We have chosen, n=25 giving $l_{c}\approx 5$. 
The saturated width is flat almost up to 5$l_{c}$, showing that $\nabla^{4}h$ 
dominates at small lengths\cite{vill2}. 
 Fig. \ref{morfw}(a) shows the morphology of the interface after 80000ML 
are grown. As predicted by the Eq.(7), the asymmetry is 
evident in the figure with $\sigma=-0.47\pm0.05$.  
. Fig. \ref {morfw}(b) shows plot of $w_{2}$ in time. 
We obtain initially $\beta$ around 0.3 that attains a steady  
value of $0.355\pm0.015$. Initial small value can be related to 
the small slope region that predicts a value of 1/3 (compare data in the  
region from 10 ML to 200 ML in the figure with the line having 
slope of $2/3$). Correspondingly h-h correlations lead to the roughness 
exponent   
$\alpha$ that increases from 0.5 to $0.65\pm0.01$. These results indicate 
that most of the morphological features of the growth with diffusion without 
detachment are captured by the current expression in Eq.(6).  
It also shows that diffusion of the 
adatoms roughens the growing surface. Diffusion bias causes additional 
effects in terms of stability or instability of growth. 
In particular, if the bias is varied from extreme -ve SE barrier 
to extreme +ve SE barrier, a stable$\rightarrow$ unstable transition 
is observed. In this transition however $h\rightarrow -h$ symmetry is
 broken asymptotically. Note that for -ve SE barrier $\nu_{2}\nabla
^{2}h$ term dominates with +ve value of $\nu_{2}$\cite{ew}, so that 
asymptotically, asymmetric term becomes irrelevant rendering  
$\sigma=0$. At exactly zero 
SE barrier, finite value of $\sigma (\approx -0.5)$
 is obtained. $\sigma$ can be  
regarded as the symmetry parameter, that changes abruptly 
 at the transition point.  
 Thus the growth transition is like 2nd order phase transition.  
Details of such a transition in computer simulations of suitable 
models will be discussed elsewhere. 

Extension of this equation to 2+1- dimensions is possible by 
similar kinetic considerations. For isotropic diffusion, same 
form as Eq.(6) is obtained with $\hat n={\bf \nabla}h/|\nabla h|$ 
and replacing length-derivative product in 
asymmetric term by $\frac{\nabla h\cdot \nabla}{(l_{c}^{-1}+
|\nabla h|)|\nabla h|}$. 
 For small slopes 
one obtains $\nabla(\nabla h)^{2}$ term\cite{ld} which seems 
to describe many experimentally observed growth roughness measurements 
from vapor\cite{krim}.   

As mentioned earlier, the present method for obtaining current from 
kinetic considerations appropriately brings out the geometrical 
dependence in growth equation. We have applied this method to one 
of the stochastic growth models proposed to capture the essential 
feature of low temperature molecular beam epitaxy (MBE) the 
DT model\cite{dt}. Based on noise reduction technique, the simulations 
of this model\cite{dtpin} confirm that,   
{\it 1)} exponent $\beta=3/8$, {\it 2)} the morphology is asymmetric 
with $\sigma \approx -0.5$ and {\it 3)} the current is tilt independent
\cite{kr1,dtpin}.  
The observed $\beta$ value of 3/8 is constant almost over 
8 orders of magnitude \cite{dtpin}. The relaxation rules for 
adatom in this model allows it to hop only when it is deposited 
at site A or B (see Fig. \ref{step}). 
Also only downward hop is allowed, if deposited at A 
and towards the step, if deposited at B. If it has choice of sites 
A,A or B,B or A,B etc. on two neighboring sites, then it will hop randomly to the 
left or right. Applying the considerations for obtaining current 
for this situation shows that the LF  
approaching sites A or B in the present set of rules is $\frac {F}
{a^{-1}+|m|a^{-1}}$, 
since the flux is affected only at large enough slope when site 
A and B overlap due to the short terrace width. 
Further, the LF is affected by relative motion of steps only when 
sites A and B differ by a lattice constant. The expression 
for the current is then, 
\begin{eqnarray} 
J_{DT}(x)&=&-\nu \frac{\partial^{2}m}{\partial{x^2}}- \frac {\hat n
Fa^{2}}{2}\nonumber\\
&&\frac {m}{1+|m|} \partial_{x}\frac {m}{(1+|m|)^{2}}
\end {eqnarray} 
From the rules it is clear that $P_{A}=P_{B}$. The growth equation 
corresponding to this current in the moving frame will be, 
\begin{equation}
\partial_{t}h=\nu\frac{\partial^{4}h}{\partial{x^4}}
+\nu_{a} \partial_{x} \frac {m}{1+|m|} \partial_{x}  
\frac {m}{(1+|m|)^{2}}+\eta
\end {equation}
where, $\nu_{a}$ accounts for various constants in the corresponding 
expression for current. 
The power counting in this equation
 leads to , $z=4$ from the first term, and $z=1+2\alpha$ for 
large slopes  corresponding to
the second term which is expected to be operative 
mainly over large inclinations. The relation obtained from 
 the second term is exactly same as 
the one obtainable from the noise term $\eta$. For $z=4$ all the 
terms are marginal. This implies that  
 $z=4$ and $\beta=3/8$. The second term is symmetry breaking. 
Hence, above equation accounts for all the observed facts mentioned 
above in the 
simulation of DT model. In 2+1-dimensions, with above rules for adatom 
relaxation, the local density of sites A and B need not be equal  
 since, fluctuations in step edges render configurations that 
show bias for sites A or B. As a result, slope dependent current 
will dominate the growth\cite{dtpin}.   
Further implications of this equation are presently under study.  

 In conclusion, we have proposed a simple method for obtaining 
current in a solid-on-solid growth in 1+1- dimension. The resultant 
equation shows that presence of diffusion alone is resposible for 
roughning of a singular surface. It induces an asymmetric term in the 
continuum equation. As a result a stable to unstable transition 
will be associated with symmetry breaking. We have also demonstrated 
the application of this method of obtaining current to a stochastic 
model referred as DT model. The corresponding continuum equation 
is shown to predict observed exponents and the nature of the 
interface profile correctly.    
 
{\it Acknowledgement:} Author acknowledges useful suggestions by 
Prof. S. Das Sarma, Univ. of Maryland, College Park, U.S.A..

\begin{references}
\bibitem{bar} See {\it e.g.} A.L. Barabasi and H.E.
Stanley, {\it Fractal
Concepts in 
Surface Growth} (Cambridge University Press, New York,
1995). 
\bibitem{kr1} J. Krug, Adv. Phys. {\bf 46}, 141 (1997).  
\bibitem{len1} J.A. Stroscio, D.T. Pierce, M.D. Stiles, A. Zangwill
, and L.M. Sander, Phys. Rev. Lett. {\bf 75}, 4245 (1995); similar 
references are obtainable from ref. 1 and 2 above. 
\bibitem{ld}Z.W. Lai and S. Das Sarma, Phys. Rev. Lett. {\bf 66 }  
, 2348 (1991).
\bibitem{vill1}J. Villain, J. Phys. I {\bf 1}, 19 (1991).     
\bibitem{vill2} P. Polity and J. Villain, Phys. Rev {\bf B} 54,
5114 (1996).
\bibitem{se} G.Ehrlich and F. Hudda, J. Chem. Phys. {\bf 44}, 
1039 (1966); R.L. Schwoebel, J. Appl. Phys. {\bf 40}, 614 (1969).      
\bibitem{tl} O.Pierre-Louis, M.R. D'Orsogna and T.L. Einstein, 
Phys. Rev. Lett. {\bf 82}, 3661 (1999); M.V. RamanMurty and 
B.H. Cooper, Phys. Rev. Lett. {\bf 83}, 352 (1999).
\bibitem{sdp}  P. Punyindu, Z. Toroczkai,and  S. Das Sarma,
Surf. Sci. Lett. {\bf 457}, L369 (2000).  
\bibitem{dt}S. Das Sarma and P. Tamborenea, Phys. Rev. Lett. {\bf 66},
 325 (1991) 
;P.Punyindu and S. DasSarma, Phys. Rev. E{\bf 57}, 
R4863 (1998).       
\bibitem{svgunpub}S.V.Ghaisas, Phys. Rev. E{\bf63}, 062601 (2001); 
S.V. Ghaisas, unpublished. 
\bibitem{ew}S.F. Edwards and   D.R.Wilkinson, Proc. R. Soc. London A{\bf 381}
,17 (1982)    
\bibitem{fam} F.Family, J. Phys. A{\bf 19}, L441 (1986); R. Lipowsky ,
J. Phys. A {\bf 18}, L585; M. Grant, Phys. Rev. B {\bf 37}, 5705 (1988). 
\bibitem{len} M.D. Johnson, C. Orme, A.W. Hunt, D. Graff, J. Sudijono, 
L.M. Sander and B.G. Orr, Phys. Rev. Lett. {\bf 72}, 116 (1994); M. Siegert 
and M. Plischke, Phys. Rev. Lett. {\bf 73}, 1517 (1994); Related work is 
found in review articles as in ref. 1 and 2 above.     
\bibitem{kim}J.M. Kim, M.A. Moore, and A.J. Brey, Phys. Rev. A{\bf 44}
, R4782 (1991). 
\bibitem{krim} J. Krim and G. Palasantzas, Int. J. Mod. Phys. B {\bf 9},
599 (1995).    
\bibitem{dtpin} S. DasSarma, P. Punyindu and Z. Toroczkai, 
Phys. Rev. E, {\it to be published};  
 P. Punyindu,
Z. Toroczkai, and S. Das Sarma, Phys. Rev {\bf B} 64, 205407 (2001).  



       


 
\end{references}
\end{document}